\documentclass[prD,showpacs,superscriptaddress,preprintnumbers,twocolumn,amsmath,nofootinbib]{revtex4}
 \usepackage[dvips,final]{graphicx}
  \usepackage{amssymb,amsmath,epsfig,bm,pifont}
      \graphicspath{{./figs/}}
\usepackage{color}
 
  \newcommand{\BluTn}[1]{\textcolor{blue}{#1}}
   \newcommand{\RedTn}[1]{\textcolor{red}{#1}}
\begin{document}
\thispagestyle{empty}
 \date{\today}
  \preprint{\hbox{RUB-TPII-04/2011}}
%\vspace*{-10mm}

\title{Pion-photon transition form factor using light-cone sum rules:\\
       theoretical results, expectations, and a global-data fit
       \footnote{Presented by the second and third authors at the 5th Joint International Hadron Structure'11 Conference,
                 Tatranska Strba (Slovak Republic),June 27--July 1, 2011.
                 }}
 \author{A.~P.~Bakulev}
  \email{bakulev@theor.jinr.ru}
   \affiliation{Bogoliubov Laboratory of Theoretical Physics, JINR,
                141980 Dubna, Russia\\}

 \author{S.~V.~Mikhailov}
  \email{mikhs@theor.jinr.ru}
   \affiliation{Bogoliubov Laboratory of Theoretical Physics, JINR,
                141980 Dubna, Russia\\}

 \author{A.~V.~Pimikov}
  \email{pimikov@theor.jinr.ru}
   \affiliation{Bogoliubov Laboratory of Theoretical Physics, JINR,
                141980 Dubna, Russia\\}

 \author{N.~G.~Stefanis}
  \email{stefanis@tp2.ruhr-uni-bochum.de}
   \affiliation{Institut f\"{u}r Theoretische Physik II,
                Ruhr-Universit\"{a}t Bochum,
                D-44780 Bochum, Germany}

\begin{abstract}
A global fit to the data from different collaborations
(CELLO, CLEO, BaBar) on the pion-photon transition form
factor is carried out using light-cone sum rules.
The analysis includes the next-to-leading QCD radiative
corrections and the twist-four contributions, while the main
next-to-next-to-leading term and the twist-six contribution
are taken into account in the form of theoretical uncertainties.
We use the information extracted from the data to investigate
the pivotal characteristics of the pion distribution amplitude.
This is done by dividing the data into two sets: one containing
all data up to 9~GeV$^2$, whereas the other incorporates also
the high-$Q^2$ tail of the BaBar data.
We find that it is not possible to accommodate into the fit
these BaBar data points with the same accuracy and conclude that
it is difficult to explain these data in the standard scheme of
OCD.
\end{abstract}
%PACS99 used: Renormalization group evolution of parameters=11.10.Hi
%             Perturbative calculations=12.38.Bx
%             Other nonperturbative calculations in QCD=12.38.Lg
%             Electromagnetic Form Factors=13.40.Gp
%\keywords{Transition form factors,
%          pion distribution amplitude,
%          higher twist,
%          light-cone sum rules,
%          collinear factorization,
%          higher-order radiative corrections,
%          renormalization group evolution}

\maketitle

\section{Form factor $\mathbf{F^{\gamma^{*}\gamma^{*}\pi}}$ in collinear QCD }
\label{sec:intro}
One of the most studied exclusive processes within QCD, based on
collinear factorization, is the pion-photon transition form factor with
both photon virtualities being sufficiently large, see \cite{BL89} for
a review.
The transition form factor is defined by the correlator of two
electromagnetic currents
\begin{eqnarray}
&&  \hspace{-10mm}\int\! d^{4}z\,e^{-iq_{1}\cdot z}
  \langle
         \pi^0 (P)\mid T\{j_\mu(z) j_\nu(0)\}\mid 0
  \rangle
\nonumber \\
&& =
  i\epsilon_{\mu\nu\alpha\beta}
  q_{1}^{\alpha} q_{2}^{\beta}
  F^{\gamma^{*}\gamma^{*}\pi}(Q^2,q^2)
 \label{eq:matrix-element}
\end{eqnarray}
%Eq (1)
with
$Q^2\equiv-q_{1}^2 >0$, $q^2\equiv -q_2^2\geq 0$,
and can be reexpressed in the form \cite{ER80}
\begin{eqnarray}
  F^{\gamma^{*}\gamma^{*}\pi}(Q^2,q^2)
&\!\!\!\!=\!\!\!\!&N\, \int_{0}^{1}\!dx\,
  T(Q^2,q^2,\mu^2_{\rm F},x)\,
\nonumber\\
& &\times \ \varphi^{(2)}_{\pi}(x,\mu^2_{\rm F})
  + O\left(\delta^2/Q^{4}\right)\,,
\label{eq:convolution}
\end{eqnarray}
%Eq (2)
by virtue of collinear factorization, assuming
that the photon momenta are sufficiently large
$Q^2, q^2 \gg m_\rho^2$.
Here $N=\sqrt{2}/3 f_\pi$, $f_\pi=132$~MeV is the pion-decay
constant, and $\delta^2$ is the twist-four coupling.
Then, the quark-gluon sub-processes, formulated in terms of
the hard-scattering amplitude of twist-two, $T$,
can be computed order-by-order of QCD perturbation theory:
$T=T_0 +a_sT_1+ a_s^2T_2+\ldots$.
The radiative corrections in next-to-leading order (NLO),
$T_1$, have been obtained in \cite{DaCh81},
the $\beta_0$--part of the contribution at the next-to-next-to-leading
order level (NNLO$_{\beta}$), encoded in the amplitude $T_2$,
i.e., $\beta_0 \cdot T_{2\beta}$, was calculated in \cite{MMP02}.

The binding  effects are separated out and absorbed
into a universal pion distribution amplitude (DA) of twist-two,
$\varphi^{(2)}_\pi(x,\mu^2)$,
defined \cite{Rad77} by the matrix element\footnote{
Gauge invariance is ensured by the longitudinal gauge link
$[z,0]=\mathcal{P}\exp \left(ig\int^z_0 A_\mu(\tau) d\tau^\mu \right)
$ along a path-ordered lightlike contour.}
\begin{eqnarray}
%&&
\langle0|\bar{q}(z)\gamma_{\nu}\gamma_{5}[z,0]q(0)|\pi(P)
   \rangle\Big|_{z^2=0}
\!\!\!&=&\! iP_{\nu} f_\pi\int_0^1 dx
      e^{ix(z\cdot P)}\nonumber\\
      &&\times \varphi^{(2)}_\pi(x,\mu^2_{\rm F}) \! \,.
\label{eq:pion-DA}
\end{eqnarray}
%Eq (3)
The variation of $\varphi^{(2)}_\pi(x,\mu^2_{\rm F})$ with the
factorization scale $\mu^2_{\rm F}$ is controlled by the
Efremov--Radyushkin--Brodsky--Lepage
(ERBL) evolution equation \cite{ER80};
moreover the Gegenbauer harmonics $\{\psi_n(x)\}$ constitute the
leading-order (LO) eigenfunctions of this equation.
Therefore, it is useful to expand the pion DA in terms of these
harmonics:
\begin{eqnarray}
  \varphi^{(2)}_{\pi}(x, \mu^2)
= \psi_{0}(x) + \sum\nolimits_{n=2,4,\ldots}
  a_{n}\left(\mu^2\right) \psi_{n}(x)\,,
\label{eq:pion-DA.Geg}
\end{eqnarray}
%Eq (4)
where $\psi_{n}(x)\!=\!6x(1-x) C^{3/2}_{n}(2x-1)$
and
$\varphi^{\rm as}(x)\!=\!\psi_{0}(x)\!=\!6x(1-x)$ is the asymptotic
pion DA~\cite{ER80}.
The nonperturbative information is contained in the coefficients
$a_{n}\left(\mu^2\right)$ with ($n\geq 2$) that have to be modeled or
extracted form the data, including evolution effects to account for
their $\mu^2$-dependence.
They are usually reconstructed from the moments
$
\langle \xi^{N} \rangle_{\pi}
\equiv
  \int_{0}^{1} dx (2x-1)^{N} \varphi_{\pi}^{(2)}(x,\mu^2)
 $
with $\langle \xi^{0} \rangle_{\pi}=1$
that can be determined by employing, e.g.,
QCD sum rules (SR)s \cite{CZ84}.
We use here the pion DA proposed before in the framework
of improved QCD SRs with nonlocal condensates (NLC-SRs) \cite{BMS01}
that yield a ``bunch''  of admissible pion DAs with two harmonics
that fix the coefficients $a_2$ and $a_4$.

\section{Light-cone sum rules for the process
$\mathbf{\gamma^*(Q^2)\gamma(q^2\simeq 0) \to \pi^0}$ }
\label{sec:LCSR}
The pion-photon transition involving two highly off-shell
photons is not easily accessible to experiment.
Experimental information is mostly available for an asymmetric
photon kinematics, with one of the photons having a virtuality
close to zero $q^2\to 0$ \cite{CELLO91,CLEO98,BaBar09}.
The calculation of this transition form factor within
perturbative QCD is a precarious step because the quasi-real photon is
emitted at large distances and has, therefore, a hadronic content
calling for the application of nonperturbative techniques.
An appropriate method is provided by light-cone sum rules (LCSRs)
\cite{Kho99} that supplements QCD perturbation
theory with a dispersion relation for
$F^{\gamma^*\gamma^*\pi}$ in the variable $q^2$, taking then
$q^2\to 0$, whereas the large variable $Q^2$ is kept fixed.
Thus, one has
\begin{equation}
  F^{\gamma^{*}\gamma^{*}\pi}\left(Q^2,q^2\right)
= N\, \int_{0}^{\infty}\!\!ds\,
  \frac{\rho\left(Q^2,s\right)}{s+q^2}\,,
\label{eq:dis-rel}
\end{equation}
%Eq (5)
with the physical spectral density $\rho(Q^2,s)$
approaching at large $s$ the perturbative one:
\begin{equation}
  \rho^{\rm PT}(Q^2,s)
=
  \frac{1}{\pi} {\rm \mathbf{Im}}\left[F^{\gamma^*\gamma^*\pi}
  \left(Q^2,-s-i\varepsilon\right)/N \right].
\label{eq:spec-dens-NLO}
\end{equation}
%Eq (6)
Using quark-hadron duality, we obtain the following LCSR
\cite{Kho99}:
\begin{eqnarray}
\!\!\!\!  Q^2 F^{\gamma^*\gamma\pi}\left(Q^2\right)
=
  \frac{Q^2}{m_{\rho}^2}
        \int_{x_{0}}^{1}\!\! \frac{dx}{x}
                        \exp\left(\frac{m_{\rho}^2-Q^2\bar{x}/x}{M^2}
                            \right)
                                   && \nonumber \\
  \times N\, \bar{\rho}(Q^2,x)
      + \int_{0}^{x_0} \frac{dx}{\bar{x}}
                     N\, \bar{\rho}(Q^2,x)&&
\label{eq:LCSR-FF}
\end{eqnarray}
%Eq (7)
with the spectral density
$\bar{\rho}(Q^2,x)=(Q^2+s)\rho^{\rm PT}(Q^2,s)$,
where $s =\bar{x}Q^2/x$ and
$x_0 = Q^2/(Q^2+s_0)$.
Note that the first term in (\ref{eq:LCSR-FF}) is
associated with the hadronic content
of a quasi-real photon at low $s\leq s_0$,
whereas the second term reproduces its point-like
behavior at the higher value $s>s_0$.
We adjust the hadronic threshold in the vector-meson
channel to the value
$s_0=1.5$~GeV$^2$, using
$m_{\rho}=0.77$~GeV
\protect\cite{PDG2010}.
We avoid to vary the Borel parameter $M^2$ in (\ref{eq:LCSR-FF})
and specify its value by virtue of
$M^2=M_{\rm 2-pt}^2/\langle{x}\rangle_{Q^2}$
entering the two-point QCD sum rule for the $\rho$-meson
with
$M_{\rm 2-pt}^2\in[0.5 \div 0.8]$~GeV$^2$,
where
$\langle{x}\rangle_{Q^2}$
denotes some average value of $x$ (at fixed $Q^2$)
in the integration region for the first integral on the
right-hand side of Eq.\ (\ref{eq:LCSR-FF}) \cite{Kho99,BMPS11}, i.e.,
$x_0(Q^2)<\langle{x}\rangle_{Q^2}<1$.

\section{Main ingredients of the LCSRs and conditions of the data analysis}
\label{sec:LCSR-data}
It is convenient to invent for each term of the harmonics $\psi_n$,
a partial spectral density according to the definition
(\ref{eq:spec-dens-NLO}) for the twist-two part \cite{MS09},
$
 \rho_n^{(i)}(Q^2,s)
 \!=\!
 \frac{\mathbf{Im}}{\pi}
                        \left[\left(T_{i}\otimes \psi_n\right)(Q^2,-s-i\varepsilon)
                        \right]
$.
The general solution for $\bar{\rho}^{(1)}_n $ in NLO
was obtained in \cite{MS09} and corrected later
(third line) in \cite{ABOP10}:
\begin{eqnarray}
\frac{1}{C_{\rm F}} \bar{\rho}^{(1)}_n\left(\frac{Q^2}{\mu^2_{\rm F}};x\right)
=
    \left\{
                    -3\left[1+v^{b}(n)\right]+\frac{\pi^2}{3}-\ln^2\left(\frac{\bar{x}}{x}\right)
                    \right.&& \nonumber \\
         \left.
           + 2v(n)
                    \ln\left(\frac{\bar{x}}{x} \frac{Q^2}{\mu^2_{\rm F}}
                       \right)
           \right\} \psi_n(x)
&&  \nonumber \\
    \!\!\!\!\!\!\!\!\!\!\!\!\!\! - 2\!
               \left[  \sum^n_{l=0,2,\ldots}\!\!\!G_{nl}\psi_l(x)
                 +v(n)\!\left(\sum^n_{l=0,1,\ldots}\!\!\!b_{n l}\psi_l(x)-3\bar{x}\!\right)\!\right],
\label{eq:spec-den-NLO} &&
\end{eqnarray}
%Eq (8)
with $v(n),v^{b}(n)$ being the eigenvalues of LO ERBL equations,
whereas $G_{nl}$ and $b_{n l}$ are calculable triangular
matrices (see \cite{MS09,ABOP10} for details).

The inclusion of the NNLO$_{\beta}$ contribution to the main partial
spectral density
$\beta_{0} \cdot \bar{\rho}^{(2\beta)}_0$,
derived from $\beta_0 \cdot T_{2\beta}$ \cite{MMP02},
was realized in \cite{MS09,BMPS11}.
It turns out that, taken together with the positive effect of a more
realistic Breit-Wigner ansatz for the meson resonance \cite{MS09}
instead of using a $\delta$-function, i.e.,
$\delta(s-m^2_{\rho})F^{\gamma^*\rho\pi}$,
as in (\ref{eq:LCSR-FF}), it is negative and about --7\%
at small $Q^2\sim 2$~GeV$^2$,
decreasing rapidly to --2.5\% at $Q^2 \geq 6$~GeV$^2$.
Here the results are expanded to include the first three harmonics.

On the other hand, the twist-six contribution to
$F^{\gamma^*\gamma\pi}$ was recently computed in \cite{ABOP10} using
$M^2 \sim 1.5$~GeV$^2$ and found to be very small.
Using instead the more moderate value
$M^2 \sim 0.75$~GeV$^2$ \cite{BMPS11}, it turns out to have
almost the same magnitude as the NNLO$_{\beta}$ term, but with the
opposite sign.

As already mentioned, we use here the BMS bunch of pion DAs
with the central point
$a_{2}^\text{BMS}(1 \text{GeV}^2)=0.20$
and
$a_{4}^\text{BMS}(1\text{GeV}^2)=-0.14$
(termed the BMS model) \cite{BMS01}.
These pion DAs have their endpoints at $x=0$ and $x=1$
suppressed---even relative to the asymptotic pion DA---and are
in good agreement with the CLEO data on the
pion-photon transition form factor as well as with the data for other
pion observables \cite{BMS02,BMS03,BMS05lat}.

\begin{figure}[h!]
 \centerline{\hspace{0mm} \includegraphics[width=0.48\textwidth]{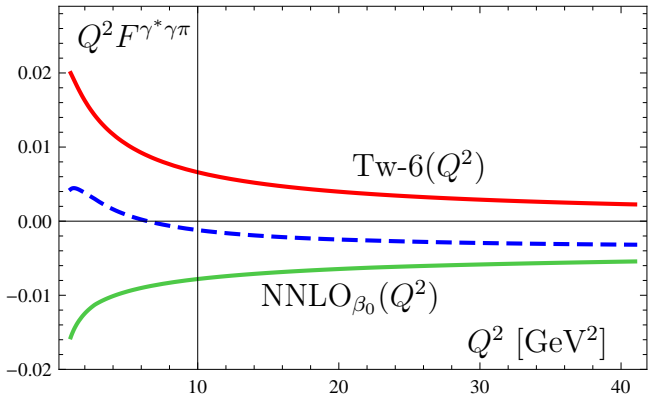} }
  \caption{\label{fig-Tw6-nnlo40.eps}
  Twist-six contribution (upper solid line in red) and
  NNLO$_\beta$ contribution (lower solid line in green), obtained with the
  BMS model, and their sum (dashed blue line).
  }
\end{figure}

The key features of our data-analysis are the following:
(i) The NLO radiative corrections in the spectral density are
included via the corrected expression (\ref{eq:spec-den-NLO}),
emphasizing that this error does not affect our previous
results in \cite{BMS02,BMS03,BMS05lat,MS09}.
The so-called default renormalization-scale setting is adopted
and, accordingly, the factorization and the renormalization scales
have been identified with the large photon virtuality $Q^2$.
(ii) The twist-four contribution is taken into account
using for the effective twist-four DA the asymptotic form
$\varphi_{\pi}^{(4)}(x,\mu^2)=
  (80/3)\,\delta^2(\mu^2)\,x^2(1-x)^2$ \cite{Kho99,BF89}.
We also admit a significant variation of the parameter
$\delta^2=0.19$~GeV$^2$
in the range 0.15~GeV$^2$ to $0.23$~GeV$^2$,
referring for details to \cite{BMS03},
and taking into account its evolution with $\mu^2$.
Using a nonasymptotic form for
$\varphi_{\pi}^{(4)}$
would not change these results significantly \cite{BMS05lat,Ag05b}.
(iii) The evolution effects of the coefficients $a_n$
are also included in NLO, employing the QCD scale parameters
$\Lambda_{\rm QCD}^{(3)}=370$~MeV and
$\Lambda_{\rm QCD}^{(4)}=304$~MeV, conforming with the NLO
estimate
$\alpha_s(M_Z^2)=0.118$ \cite{PDG2010}.
(iv) \label{page4}
The NNLO$_{\beta}$ radiative correction to the
LCSR form factor \cite{MS09,BMPS11}, Eq.\ (\ref{eq:LCSR-FF}),
is incorporated together with the twist-six term,
computed in \cite{ABOP10}, in terms of theoretical uncertainties.
To be precise, the calculation of the NNLO$_{\beta}$ term
involves only the convolution of the hard-scattering amplitude
$T_\beta$ with the DA based on the three lowest harmonics.
This treatment makes sense due to the fact that for the average value
of $M^2(Q^2)\sim 0.75$~GeV$^2$, these two contributions
almost mutually cancel and the net result is small---see
Fig.\ref{fig-Tw6-nnlo40.eps}.

%%%%%%%%%%%%%%%%%%%%%%%%%%%%%%%%%%%%%%%%%%%%%%%%%%%%%%%%%%%%%%%%%%%%%%%
%%%%                             Figure 1                          %%%%
%%%%%%%%%%%%%%%%%%%%%%%%%%%%%%%%%%%%%%%%%%%%%%%%%%%%%%%%%%%%%%%%%%%%%%%
%%%%%%%%%%%%%%%%%%%%%%%%%%%%%%%%%%%%%%%%%%%%%%%%%%%%%%%%%%%%%%%%%%%%%%%
In fact, it decreases with $Q^2$ from $+0.004$ at
$Q^2=1$~GeV$^2$---where the twist-six term dominates---down
to $-0.003$ at $Q^2=40$~GeV$^2$---where the NNLO$_{\beta}$ correction
starts prevailing.
This particular behavior applies only to the moderate value
of the Borel parameter $M^2=0.75$~GeV$^2$ \cite{BMPS11}, while
for the larger value $M^2=1.5\pm0.5$~GeV$^2$, used in \cite{ABOP10},
the twist-six term would be much
smaller and the net result would be everywhere negative and almost
constant: $\approx-0.004$.

\section{Data Analysis}
\label{sec:datA}
Here we overview our fit procedure of all available experimental
data on the pion-photon transition form factor
$F^{\gamma^*\gamma\pi}$,
within the framework of LCSRs,
as worked out in \cite{BMPS11}.
The main goal of the fit is to extract the pion DA---the main
low-energy pion characteristic---best compatible with all the data.
This is done fitting the form factor by varying the pion DA
in terms of the Gegenbauer coefficients $a_n$.
To reveal the particular role of the new high-$Q^2$ BaBar data in the
fit, we perform our analysis utilizing two different data sets.
The first set (set-1) contains all available data from CELLO
\cite{CELLO91}, CLEO \cite{CLEO98}, and BaBar \cite{BaBar09} that
belong to the $Q^2$-window $[1\div 9]$~Gev$^2$.
The second set (set-2) comprises all data in the range
$[1\div 40]$~GeV$^2$.
First, we define the optimal number of Gegenbauer harmonics necessary
to model the pion DA.
Second, we determine the fiducial regions of the
corresponding coefficients $a_n$.
Third, we relate these regions with the pion DA and its
characteristics: profiles, derivatives at the origin, and its moments.
Finally, we confront the obtained results with the data of set-1 and
set-2.

\subsection{How many harmonics should be taken into account?}
\begin{figure}[h!]
 \centerline{\hspace{0mm}\includegraphics[width=0.48\textwidth]{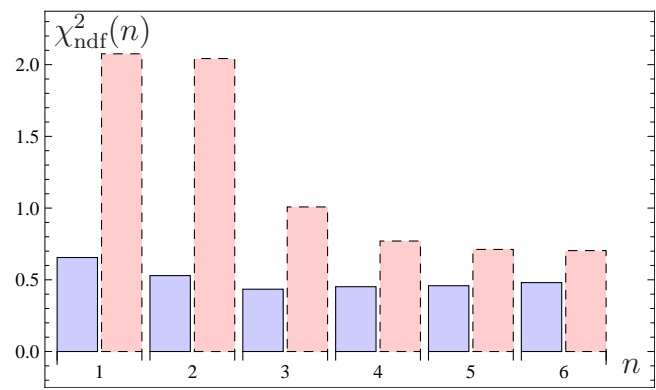}}
  \caption{\label{fig-chi-n.eps}
   Dependence of the goodness of fit $\chi_\text{ndf}^2\equiv\chi^2/{\rm ndf}$
   (ndf~$=$~number of degrees of freedom) on the number $n$ of
   Gegenbauer harmonics shown as histograms: set-1---solid (blue) bars;
   set-2---higher dashed (red) bars.}
\end{figure}

To answer this question, we confront the dependence
of the fit quality on the number of the parameters of the involved
harmonics and the associated statistical errors.
The statistical errors in the parameter determination increase
with their number for statistical reasons, while the
$\chi_\text{ndf}^2$ initially decreases.
Therefore, in order to achieve an acceptable compromise, one should
use the lowest acceptable number of harmonics.
The dependence of the goodness of fit, $\chi_\text{ndf}^2$, on the
number $n$ of the involved harmonics for the two data sets, is
presented in Fig. \ref{fig-chi-n.eps}.
%%%%%%%%%%%%%%%%%%%%%%%%%%%%%%%%%%%%%%%%%%%%%%%%%%%%%%%%%%%%%%%%%%%%%%%
%%%%%%%%%%%%%%%%%%%%%%%%%%%%%%%%%%%%%%%%%%%%%%%%%%%%%%%%%%%%%%%%%%%%%%%
%%%%                             Figure 2                          %%%%
%%%%%%%%%%%%%%%%%%%%%%%%%%%%%%%%%%%%%%%%%%%%%%%%%%%%%%%%%%%%%%%%%%%%%%%
%%%%%%%%%%%%%%%%%%%%%%%%%%%%%%%%%%%%%%%%%%%%%%%%%%%%%%%%%%%%%%%%%%%%%%%
The goodness of fit for set-1 is only slightly decreasing with $n$
and remains almost stable after $n=3$.
Thus, 2 to 3 parameters are actually enough to describe all
data in this region with
$\chi_\text{ndf}^2 \approx 0.5$.
In contrast, the data description of set-2 is only possible with a
$\chi_\text{ndf}^2$ value 2 or 3 times larger---even if we
include more harmonics.
To fit all the data, we are forced to consider
\textit{at least} 3 parameters with $\chi_\text{ndf}^2 \approx 1$.
To have an even better description with a goodness of fit approximately
equal to 0.8, we have to employ 4 parameters.
Further increase of the number $n$ will not provide any improvement.

\begin{figure}[h!]
 \centerline{\hspace{0mm}\includegraphics[width=0.48\textwidth]{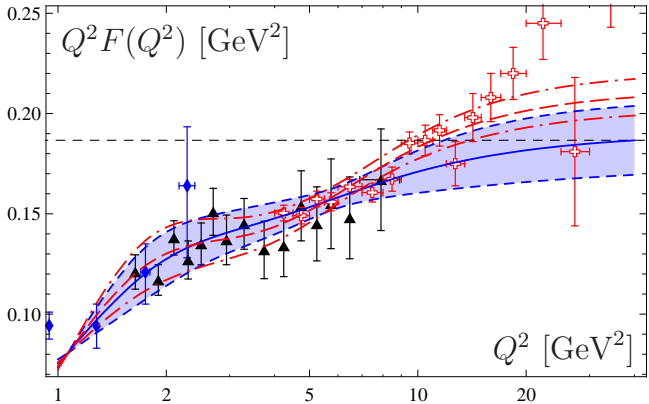}}
  \caption{\label{fig.pionFF.fit}
    Best-fit curves to the experimental data for the transition
    form factor in the framework of LCSRs:
    solid (blue) line---best-fit curve of set-1;
    strip bounded by dashed (blue) lines---68\% CL region;
    long dashed (red) line---best-fit curve of set-2;
    strip bounded by dashed dotted (red) lines---68\% CL region.
    Error bars show the sum of the statistical errors and the
    twist-four uncertainties.
    The experimental data are taken from the
    CELLO \cite{CELLO91} (diamonds),
    CLEO \cite{CLEO98} (triangles),
    and BaBar \cite{BaBar09} (open crosses) experiments.
         }
\end{figure}

However, for the sake of comparison of the results, one should use
the same fit model of pion DA, which means that the most appropriate
number of harmonics may be fixed to 3.
Best-fit curves for both data sets are shown
in Fig. \ref{fig.pionFF.fit} as a bunch of form-factor predictions with
errors stemming from the sum of the statistical error and the twist-four
uncertainties.
At high values of the momentum transfer, the fit curve of the
set-2 data---long dashed (red) line---exceeds the 68\% CL
(confidential level) region of the set-1 data fit---solid (blue) line.
This indicates that in the framework of LCSRs, the new BaBar data above
9 GeV$^2$ deviate from the low-$Q^2$ data at the level of a 1$\sigma$
deviation and more.
%%%%%%%%%%%%%%%%%%%%%%%%%%%%%%%%%%%%%%%%%%%%%%%%%%%%%%%%%%%%%%%%%%%%%%%
%%%%%%%%%%%%%%%%%%%%%%%%%%%%%%%%%%%%%%%%%%%%%%%%%%%%%%%%%%%%%%%%%%%%%%%
%%%%                             Figure 3                          %%%%
%%%%%%%%%%%%%%%%%%%%%%%%%%%%%%%%%%%%%%%%%%%%%%%%%%%%%%%%%%%%%%%%%%%%%%%
%%%%%%%%%%%%%%%%%%%%%%%%%%%%%%%%%%%%%%%%%%%%%%%%%%%%%%%%%%%%%%%%%%%%%%%
%%%%%%%%%%%%%%%%%%%%%%%%%%%%%%%%%%%%%%%%%%%%%%%%%%%%%%%%%%%%%%%%%%%%%%%
\subsection{Data analysis vs pion DA models}
Performing the data analysis, we obtain the best-fit values of
the pion DA in the 68\% CL region for a number of harmonics $n=2\div3$.
The 3D graphics of the confidential regions for the
3 harmonics analysis were presented in our recent
work in \cite{BMPS11}, whereas the best-fit values together
with the statistical errors and the twist-four uncertainties are given
in Table~\ref{table:chi} below.
We compare there our fit results with various pion DA models in terms
of the goodness of fit $\chi_\text{ndf}^2$ for the two analyzed sets of
experimental data.
From the first two lines of Table~\ref{table:chi}, we infer that the
inclusion of the new high-$Q^2$ BaBar data affect only the value of the
parameter $a_6$, while $a_2$ and $a_4$ do not change significantly.
Moreover, the good description of the experimental data up to
9~GeV$^2$ becomes appreciably  worse after the inclusion of the
high-$Q^2$ tail of the BaBar data.
The BMS pion DA stands out in the sense that it provides
the best fit for set-1, while all other models cannot reproduce these
data good enough.

\begin{figure}[t!]
\centerline{\hspace{0mm}\includegraphics[width=0.48\textwidth]{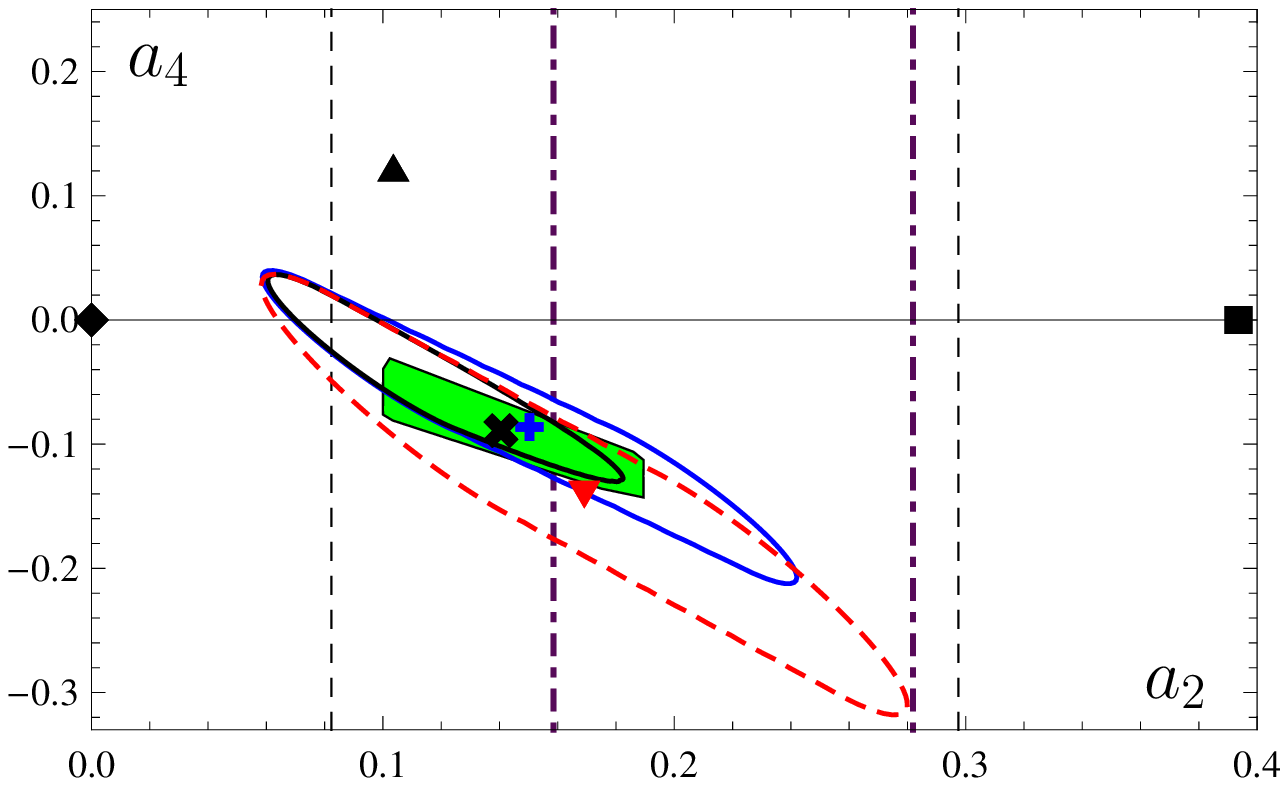}}
\centerline{\hspace{0mm}\includegraphics[width=0.48\textwidth]{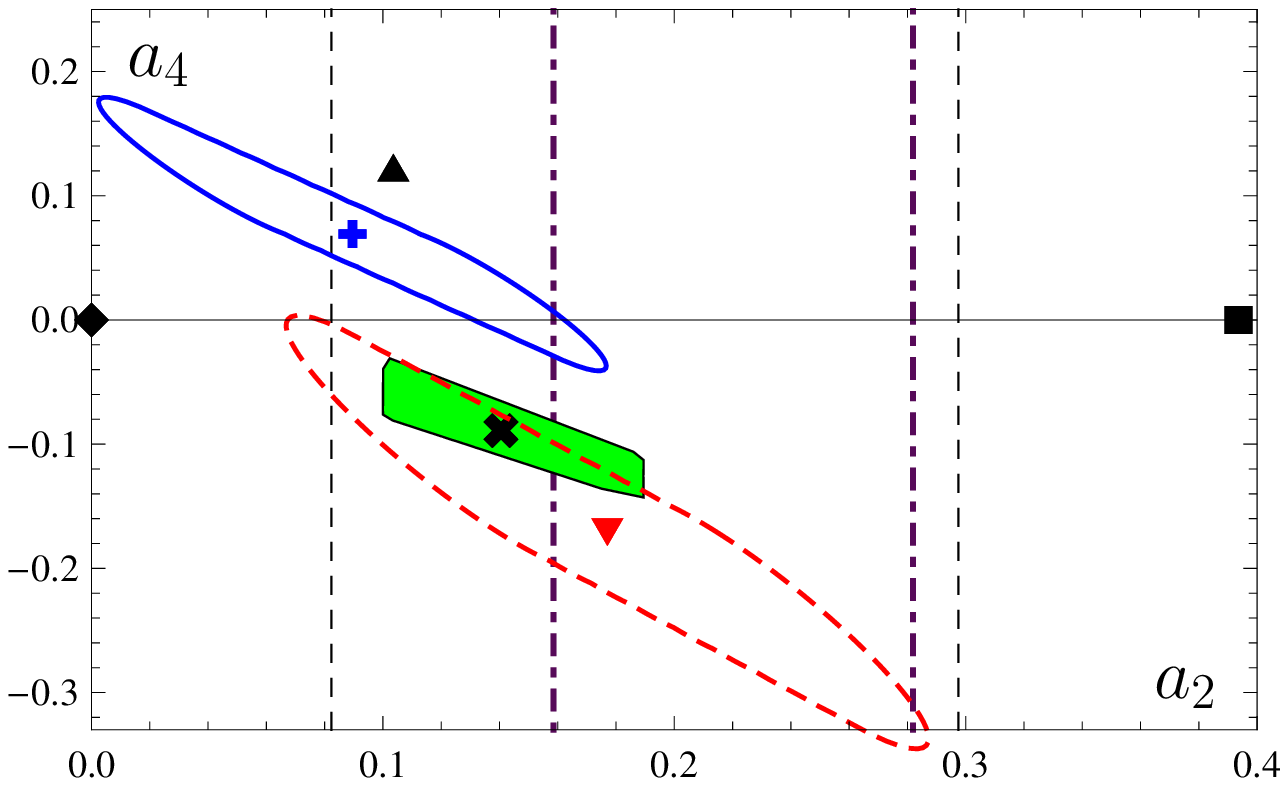}}
  \caption{(color online). Distorted $1\sigma$ error ellipses
  for set-1 (upper panel) and set-2 (lower panel) from various experiments
  \protect\cite{CELLO91,CLEO98,BaBar09,BaBar11-BMS}
  using different data-analysis procedures.
  These ellipses result from merging together the ellipses
  associated with different values of the twist-four parameter in the range
  $\delta^2=0.15 \div 0.23$~GeV$^2$.
  The slanted shaded (green) rectangle encloses the area of $a_2$ and
  $a_4$ values determined by NLC-SRs \protect\cite{BMS01}, with the
  BMS pion DA being marked by {\footnotesize\ding{54}}.
  The middle points of the ellipses (\BluTn{\footnotesize\ding{58}} and
  \RedTn{\footnotesize\ding{116}}), the asymptotic DA ({\footnotesize\ding{117}}),
  the CZ DA ({\footnotesize\ding{110}}), and Model III from
  \protect\cite{ABOP10} ({\footnotesize\ding{115}}) are also marked.
  The range of values of $a_2$, restricted by lattice simulations, are indicated
  by vertical lines: \cite{Lat06}---dashed lines;
  \cite{Lat10}---dashed-dotted (blue) lines.
  All results are shown at the scale $\mu_\text{SY}^2=(2.4$~GeV$)^2$,
   whereas
  the treatment of the Borel parameter $M^2(Q^2)$ is explained in the text.
  Graphics taken from \protect\cite{BMPS11}.
 \protect\label{fig:data-ellipses}}
\end{figure}
%\end{widetext}
%%%%%%%%%%%%%%%%%%%%%%%%%%%%%%%%%%%%%%%%%%%%%%%%%%%%%%%%%%%%%%%%%%%%%%%

It is worth remarking that one of the models, obtained from fitting
the experimental data within the Modified Factorization Scheme (MFS),
has the value $\chi_\text{ndf}^2=4.4$ in contrast to the result
$\approx 1.1$ obtained in \cite{Kro10sud}.
This discrepancy indicates that using the same pion DA in the
framework of LCSRs and the MFS may lead for the same observable to
incompatible results---a theoretical bias.

\renewcommand{\tabcolsep}{0.85pc} % enlarge column spacing
\renewcommand{\arraystretch}{1.3} % enlarge line spacing
\begin{table*}[h!t]
\caption{
 Measures of goodness of fit of selected pion DA models (first column)
 with associated coefficients $a_n$ (second column), used in the
 calculation of the pion-photon transition form factor by means
 of LCSRs.
 Note that the coefficients $a_n$ are strongly correlated
 and the errors of $a_n$ represent the maximal variation in the range
 of the $1\sigma$-region.
 The last two columns show the values of
 $\chi_\text{ndf}^2$
 for the data in set-1 and for the whole set
 of the data, (set-2), respectively.
 All values of the coefficients $a_n$  are given at the scale
 $\mu_{\text{SY}}=2.4$~GeV \protect\cite{SY99}.
 }
\label{table:chi}
 \centerline{
\begin{tabular}{lccc}\hline
 Model/Fit                      &  $(a_2, a_4, \ldots )_{\mu^2=\mu_\text{SY}^2}$
                                                       & $\chi^2_\text{ndf},~[1-9]$~GeV$^2$ & $\chi^2_\text{ndf},~[1-40]$~GeV$^2$
                                                       \\ \hline
%                                &                      &                                    &       \\
    3D fit, $[1-9]$~GeV$^2$     & $(0.17\pm 0.11,-0.14\pm 0.18,0.12\pm 0.17)$
                                                       & $0.4$               & $-$                     \\%\hline
    3D fit, $[1-40]$~GeV$^2$    & $(0.18\pm 0.11,-0.17\pm 0.17,0.31\pm 0.12)$
                                                       & $-$                 & $1.0$                     \\%\hline
    NLC-SRs, BMS~\cite{BMS01}& $(0.141,-0.089)$     & $0.5$               & $3.1$                      \\%\hline
    Model I \protect\cite{ABOP10}& $(0.084,0.137,0.088)$& $\geq 2.8$          & $\geq 2.4$                  \\%\hline
    Modif. fact.~\cite{Kro10sud}& $(0.21, 0.009)$      & $3.8$               & $4.4$                        \\%\hline
    AdS/QCD,~\cite{BT07}        & $(0.15, 0.06, 0.03,\ldots)$
                                                       & $2.3$               & $2.8$                         \\%\hline
    CZ \cite{CZ84}              & $(0.394,0)$          & $32.3$              & $25.5$                         \\%\hline
    Asympt.                     & $(0,0)$              & $4.7$               & $7.9$                           \\\hline
\end{tabular}
}
\end{table*}
Below, we consider in detail the results of the 2D analysis in the
$(a_2,a_4)$ plane, presented in Fig.\ \ref{fig:data-ellipses},
with the upper panel showing the results for set-1, whereas the
lower panel presents those for set-2.
To this end, we calculate the $1\sigma$ error ellipses%
\footnote{%%%
We denote by a $1\sigma$ ellipse (ellipsoid) a $68.27\%$
confidence-level boundary.}
by allowing the parameter $\delta^2$ to vary by $20\%$ around
the value 0.19~GeV$^2$.
The obtained error ellipses are then unified into a single
(distorted) $1\sigma$ ellipse shown in Fig.\
\ref{fig:data-ellipses}.
To be specific, we consider the following cases:
(i) The result of combining
the projections on the plane $(a_2,a_4)$ of the 3D (3 parameter)
data analysis is represented by the largest ellipse---dashed (red)
line with the middle point \RedTn{\footnotesize\ding{116}}.
(ii) The analogous result of the 2D (2 parameter) data analysis
in terms of $a_2$ and $a_4$ is shown by the smaller ellipse
(solid blue line) with the middle point \BluTn{\footnotesize\ding{58}}
having the coordinates $(0.15,-0.09)$
and $\chi^2_\text{ndf} \approx 0.5$,
that almost coincides with the middle point {\footnotesize\ding{54}}
of the parameter area determined by NLC-SRs \cite{BMS01}.
(iii) The combination of the intersections with the
$(a_2,a_4)$ plane of all 3D ellipsoids generated by the variation
around the central value of $\delta^2$ give rise to the smallest
ellipse (thick line), entirely enclosed by the previous one.

For convenience, the locations in the $(a_2,a_4)$ plane of some
characteristic pion DAs are also indicated in
Fig.\ \ref{fig:data-ellipses}.
These are the asymptotic DA ({\footnotesize\ding{117}}),
the CZ model ({\footnotesize\ding{110}}),
and the projection of Model III from \cite{ABOP10}
({\footnotesize\ding{115}}).
Note that the slanted (green) rectangle, containing those
values of $a_2$ and $a_4$ that have been determined by
NLC-SRs \cite{BMS01}, is practically within both larger error
ellipses and also overlapping with the smallest one.
Moreover, the BMS model DA {\footnotesize\ding{54}}
stands out by lying inside of all $1\sigma$ error ellipses.
Thus, the theoretical predictions obtained from the 2D and the 3D
data analyses conform with each other and agree at the level
of $\chi^2_\text{ndf}\leq0.5$ with the results
obtained from NLC-SRs \cite{BMS01}.
The calculated $1\sigma$ error ellipses comply rather good with
the boundaries for $a_2$ extracted from two independent
lattice simulations.
The vertical dashed lines denote in both panels the older
estimate from \cite{Lat06}, while the very recent
constraints from Ref.\ \cite{Lat10} are represented by the
dashed-dotted (blue) vertical lines.

From the lower panel of Fig.\ \ref{fig:data-ellipses} it
becomes evident that the situation changes significantly when including
in the analysis the high-$Q^2$ tail of the
BaBar data \cite{BaBar09}.
Indeed, using the same designations as in the upper panel,
we display the analogous unified error ellipses and
observe that the error ellipsoid has \textit{no intersection}
with the $(a_2,a_4)$ plane, whereas the composed error ellipse
resulting from the 2D analysis (solid blue line) deviates from the
region of negative values of $a_4$ and moves inside its positive
domain.
At the same time, the fit quality deteriorates yielding
$\chi^2_\text{ndf}\approx 2$,
as opposed to the value $\chi^2_\text{ndf}\approx 0.5$
determined for set-1 of the data.
As regards the unified $1\sigma$ error ellipse of
the 3D projections on the $(a_2,a_4)$ plane
(larger dashed red ellipse),
its position remains unaffected, still enclosing most of the area of
the $a_2$, $a_4$ values computed with NLC-SRs---shaded (green)
rectangle.

  \begin{figure*}[t]
 \centerline{
    \includegraphics[width=0.48\textwidth]{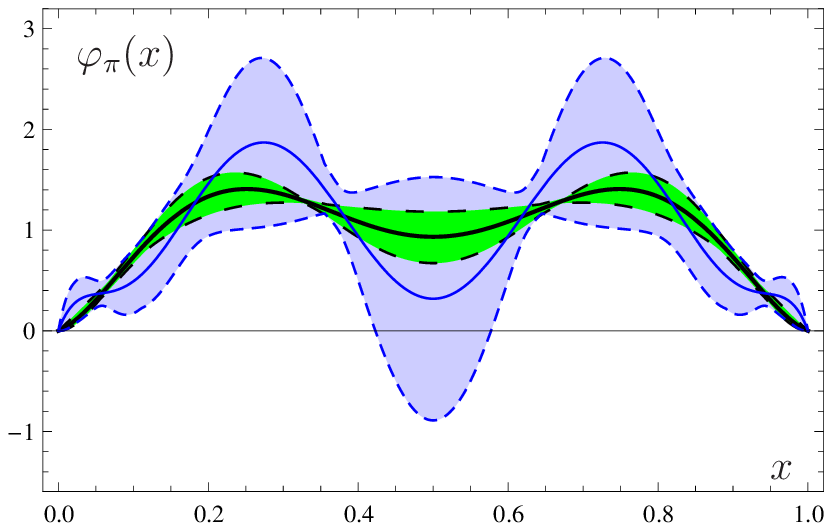}
~~~~~\includegraphics[width=0.48\textwidth]{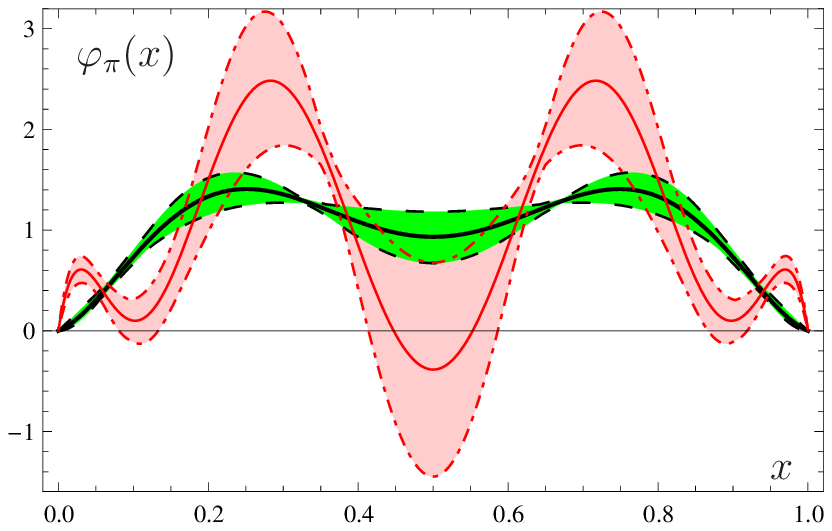}}
  \caption{\label{fig-piDA}
  Left. Comparison of the BMS pion DA bunch (shaded strip in green color)
  and of the BMS model (black solid line inside this strip) with the 3D
  fit to the experimental data on the pion-photon transition form factor.
  The solid blue line denotes the best-fit pion DA sample obtained from
  the analysis of set-1, with the dashed lines indicating the sum of the
  statistical errors of the fit and the twist-four uncertainties.
  Right. Analogous results obtained with set-2.
   }
\end{figure*}
%%%%%%%%%%%%%%%%%%%%%%%%%%%%%%%%%%%%%%%%%%%%%%%%%%%%%%%%%%%%%%%%%%%%%%%

The high quality of the data fit parallels the lattice findings,
with the 3D error ellipse being almost entirely inside the boundaries
from \cite{Lat06} (dashed vertical lines), while it also overlaps
for the larger values of $a_2$ with the range of values computed in
\cite{Lat10} (dashed-dotted vertical lines).
In contrast, the ellipse from the 2D analysis agrees very roughly
with the small $a_2$ window of \cite{Lat10}, sharing also only a small
common area with the low end of the $a_2$ region found in \cite{Lat06}.
Obviously, no agreement between the 2D and the 3D analysis is found.
This discrepancy is also reflected in the values of the
$\chi$-criterion of the 2D fit model $\chi^2_\text{ndf}\approx 2$ and
that for the 3D model which turns out smaller by a factor of 2:
$\chi^2_\text{ndf}\approx 1$.
This means that a pion DA, based only on 2 harmonics, is not
sufficient to describe all the data on the pion-photon transition
form factor.
This deviating behavior of the results, associated with the fits for
set-1 and set-2, shows up for a larger number of degrees of freedom,
i.e., when including into the data analysis the next higher
harmonics $\psi_n$ with $n=6,8,10$.
But, for the case of the set-1 fit, this expansion does not improve
any further the value $\chi^2_\text{ndf}\leq 0.5$---this
remains approximately stable.
Moreover, the $1\sigma$-admissible regions in 2D, 3D, or 4D
parameterizations appear to be each embedded inside the other.
In contrast, fitting the set-2 data, these new degrees of freedom
lead to a decrease of $\chi^2_\text{ndf}$, while the corresponding
$1\sigma$-regions in the 2D, 3D, or 4D space,
either do not overlap at all or intersect only marginally.

It becomes obvious from Fig.\ \ref{fig:data-ellipses} that Model III
({\footnotesize\ding{115}}) from \cite{ABOP10} has a projection on the
$(a_2,a_4)$ plane that lies outside of all considered $1\sigma$ error
ellipses of the data.
However, selecting for the Borel parameter
the value $M^2=1.5$~GeV$^2$, as in \cite{ABOP10},
the agreement of this model with the data improves to the level of
$\chi^2_\text{ndf}\gtrsim 1.5$.

\subsection{Pion DA  characteristics}
The confidential region of the coefficients $\{a_n\}$, obtained above,
can be linked to any other characteristic of pion DA.
The profiles of the pion DA $\varphi_\pi(x)$, extracted in the 3D fit
procedure, are shown in Fig. \ref{fig-piDA}: left panel---set-1;
right panel---set-2.
The BMS bunch (shaded green strip) and the BMS DA model
(black solid line) are also shown in both cases.
The inclusion into the data fit of the high-$Q^2$ BaBar tail,
causes a modification of the shape of the pion DA---see Fig.
\ref{fig-piDA}---giving support to our previous observation that the
BMS bunch is within the error range of the set-1 fit (left panel),
while the best fit to set-2 differs considerably (right panel).
In addition, the pion DA becomes endpoint enhanced, as opposed to the
endpoint-suppressed BMS pion DA.
The endpoint behavior can be characterized by its slope at the origin
given by the derivative $\varphi_\pi'(0)$ or, more adequately, by
the so-called ``integral derivative'' $D^{(2)}\varphi_\pi(x)$,
introduced in~\cite{MPS10}.
The integral derivative is the average derivative $\varphi'_\pi(x)$
defined by
$$D^{(2)}\varphi(x)=\frac{1}{x}\int_0^x\frac{\varphi(y)}{y}dy$$
with the important property
$\lim\limits_{x\to 0} D^{(2)}\varphi(x)=\varphi'_\pi(0).$
\begin{table}[h!]
\caption{
Comparison of the pion DA characteristics for the data of set-1
and set-2.
\label{tab:sets}}
\centerline{
\begin{tabular}{lcc}                                                           \hline
data set                   & $[1-9]$~GeV$^2$        & $[1-40]$~GeV$^2$         \\\hline%\hline
$\varphi'_\pi(0)$          & $20.2\pm 19.8\pm 1.1$  & $48.5\pm 11.4\pm 0.4$    \\%\hline
$D^{(2)}\varphi_\pi(0.4)$  &  $6.6\pm 1.1\pm 0.4$   & $8.1\pm 0.7\pm 0.3$      \\%\hline
BMS DA                     & Agreement              & No                       \\%\hline
$n$                        &   2, 3                 & 3, 4                     \\%\hline
$\chi^2_\text{ndf}$        & 0.53,\,\,\, 0.44       & $1.0,\,\,\, 0.77$        \\\hline
\end{tabular}}
\end{table}
\\
Using a 3D confidential bound on the Gegenbauer coefficients, we get
the values of the derivatives
$\varphi'_\pi(0)$ and $D^{(2)}\varphi(0.4)$, supplied in
Table \ref{tab:sets} for both data sets.
These characteristics are shown together with the theoretical errors,
the first being statistical and the second stemming from the twist-four
uncertainty.
We observe that these characteristics can clearly differentiate the
pion DAs generated from set-1 and set-2.
\subsection{Combining Lattice constraints with experimental data}
%%%%%%%%%%%%%%%%%%%%%%%%%%%%%%%%%%%%%%%%%%%%%%%%%%%%%%%%%%%%%%%%%%%%%%% FIGURE 5
\begin{figure}[t!]\vspace*{-3mm}
\centerline{\hspace{0mm}\includegraphics[width=0.48\textwidth]{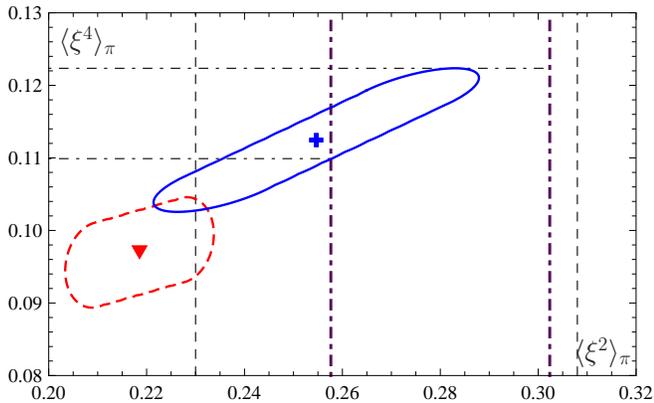}}
 \vspace*{-3mm}
 \caption{(color online). Predictions for the moments
  $\langle \xi^2\rangle_\pi$ and
  $\langle \xi^4\rangle_\pi$
  at the lattice scale $\mu^2_\text{Lat}=4$~GeV$^2$.
  The solid (blue) ellipse corresponds to our choice of $M^2$,
  whereas the dashed (red) one results when using $M^2=1.5$~GeV$^2$.
  The vertical lines show the range of values
  computed on the lattice:
  dashed line---\protect\cite{Lat06};
  dashed-dotted (violet) line---\protect\cite{Lat10}.
 \protect\label{fig:Moments}}
\end{figure}
%%%%%%%%%%%%%%%%%%%%%%%%%%%%%%%%%%%%%%%%%%%%%%%%%%%%%%%%%%%%%%%%%%%%%%%
The possibility to extract information on the moment
$\langle \xi^4 \rangle_\pi$ of the pion DA by combining lattice
constraints with the experimental data was first pointed out in
\cite{Ste08} and the following range of values was extracted from the
$1\sigma$ error ellipse of the CLEO data \cite{CLEO98} in conjunction
with the lattice constraints for
$\langle \xi^2 \rangle_\pi$ from \cite{Lat07}:
$\langle \xi^4 \rangle_\pi \in [0.095 \div 0.134]$ at
$\mu^2_\text{Lat}=4$~GeV$^2$ and for $M^2=0.7$~GeV$^2$.
This procedure was refined in \cite{BMPS11} in the following way:
first, we expanded the result of the 2D analysis to the
$(\langle \xi^2 \rangle_\pi,\langle \xi^4 \rangle_\pi)$ moments.
Then, we determined the intersection of the confidential region
(the area enclosed by a solid blue line) in Fig.\ \ref{fig:Moments}
for set-1 (cf.\ Fig.\ \ref{fig:data-ellipses})
using the constraints from \cite{Lat06} and \cite{Lat10}.
The intersection of these constraints, evaluated at the typical
lattice scale $\mu^2_\text{Lat}=4$~GeV$^2$, and the experimental data
leads to the following moment results, respectively,
(i) $\langle \xi^2 \rangle_\pi \in [0.23 \div 0.29]$
and
$\langle \xi^4 \rangle_\pi \in [0.102 \div 0.122]$,
(ii)
$\langle \xi^2 \rangle_\pi \in [0.26 \div 0.29]$
and
$\langle \xi^4 \rangle_\pi \in [0.11 \div 0.122]$.
These common validity ranges were extracted using a
$Q^2$-dependent
Borel parameter---like everywhere in our analysis here and in
\cite{BMPS11}.
On the other hand, the value $M^2=1.5$~GeV$^2$ \cite{ABOP10},
yields only a small intersection of the validity region extracted
from set-1 (shown in Fig.\ \ref{fig:Moments} by the dashed red line)
with the lattice constraints of~\cite{Lat06}.
This restricts the common region of validity to the value
$\langle \xi^4 \rangle_\pi \simeq0.1$,
whereas there is no intersection at all with the lattice estimates
from \cite{Lat10}.
This obvious sensitivity of $\langle \xi^2 \rangle_\pi$ on
the choice of the particular value of the Borel parameter
$M^2$ gives additional support to our choice of the value of the
Borel parameter.

\section{Conclusions}
We have presented here a global fit to the data on the
pion-photon transition form factor, discussing further our recent
analysis in \cite{BMPS11}.
To get a precise measure of the influence of the high $Q^2$ BaBar
data on the form factor and the pion DA, we divided the experimental
data in two different sets with respect to $Q^2$.
Set 1 contains all data in the range $[1\div 9]$~GeV$^2$,
whereas the second set comprises all data in the regime
covered by BaBar, i.e., $[1\div 40]$~GeV$^2$.
As a result, we obtained the confidential regions of different
characteristics of the pion DA (Gegenbauer coefficients,
derivatives of $\varphi_\pi(x)$ at $x=0$, and its
moments) by fitting the experimental data within the framework of LCSRs.
The predictions obtained from the CELLO, CLEO, and the BaBar data up to
9~GeV$^2$ are in good agreement with the previous fits, based only the
CLEO data \cite{SY99,BMS02,BMS03,BMS05lat,MS09},
giving preference to an endpoint-suppressed pion DA \cite{BMS01}.
Beyond 9~GeV$^2$, the best fit requires a sizeable coefficient $a_6$
that inevitably leads to an endpoint-enhanced pion DA.
The data analysis tells us that the inclusion of the high-$Q^2$ tail
of the BaBar data affects mainly the Gegenbauer coefficient $a_6$,
while $a_2$ and $a_4$ change only insignificantly.
The good description of the experimental data up to 9~GeV$^2$ using
LCSRs becomes considerably less accurate after the inclusion of the high-$Q^2$
data but yields an acceptable value of $\chi^2_\text{ndf} \approx 1$.
This effect has been discussed before in \cite{MS09Trento} at a qualitative level.
The results obtained with the inclusion of the high-$Q^2$ tail of the
BaBar data indicate a possible discrepancy between the result of
the BaBar experiment and the method of LCSRs.
Indeed, the high-$Q^2$ BaBar data require a pion DA with a sizeable
number of higher Gegenbauer coefficients $a_n$, or alternative
theoretical schemes outside the standard QCD factorization approach,
see, e.g., \cite{Rad09,Pol09,Dor09,KOT10plb,SZ11,WH10}.
Similar conclusions were also drawn in \cite{RRBGGT10} using
Dyson--Schwinger equations and in the recent works \cite{BCT11},
based on AdS/QCD.

\section{Acknowledgments}
%%%%%%%%%%%%%%%%%%%%%%%%%%%%%%%%%%%%%%%%%%%%%%%%%%%%%%%%%%%%%%%%%%%%%%%
We would like to thank Simon Eidelman, Andrei Kataev, and Dmitri Naumov
for stimulating discussions and useful remarks.
A. P. wishes to thank the Ministry of Education and Science of
the Russian Federation
(``Development of Scientific Potential in Higher Schools''
 projects No.\ 2.2.1.1/12360 and No.\ 2.1.1/10683).
This work was supported in part by the Heisenberg--Landau Program under
Grant 2011, the Russian Foundation for Fundamental
Research (Grant No.\ 11-01-00182), and the BRFBR--JINR Cooperation
Program under contract No.\ F10D-002.

\end{document}